\documentclass{article}
\usepackage{graphicx} 
\usepackage{graphics}
\usepackage{amsthm}
\usepackage{amsmath}
\usepackage{mathrsfs}
\usepackage{graphicx}
\usepackage{floatrow}
\usepackage{caption}
\usepackage{subcaption}
\usepackage{soul}
\usepackage{mathtools}
\usepackage[margin=2.5cm]{geometry}
\DeclarePairedDelimiter\bra{\langle}{\rvert}
\DeclarePairedDelimiter\ket{\lvert}{\rangle}
\DeclarePairedDelimiterX\braket[2]{\langle}{\rangle}{#1\,\delimsize\vert\,\mathopen{}#2}
\usepackage{mathrsfs}
\usepackage{authblk}

\title{Jones Matrix Reconstruction with Undetected Photons}

\author[1]{Gaytri Arya}
\author[1]{Paolo Bianchini}
\author[1,2]{Alberto Diaspro}

\affil[1]{Nanoscopy, Istituto Italiano di Tecnologia, Genoa, Italy}
\affil[2]{DIFILAB Department of Physics, University of Genoa, Genoa, Italy}

\begin{document}

\maketitle
\begin{abstract}
We present a theoretical polarization tomography scheme within the QIUP framework that directly extracts key sample parameters from measured interference visibilities and phase shifts. This approach establishes constant bindings for transmission amplitudes and coherence terms, thereby reducing the number of unknowns required to reconstruct the sample’s Jones matrix. Our method is resource-efficient and well-suited for quantum imaging where probing and detection occur at different wavelengths, enabling robust, non-invasive characterization of complex samples. 
\end{abstract}

\section{Introduction}
Quantum imaging with undetected photons (QIUP) has rapidly evolved as a powerful technique in quantum optics, leveraging the phenomenon of induced coherence without induced emission \cite{zou1991induced,lemos2014quantum,gilaberte2019perspectives}. The core principle relies on the indistinguishability of idler photons produced by two spatially separated sources, enabling imaging at infrared (IR) wavelengths while detection occurs at visible wavelengths \cite{kviatkovsky2022mid}. This approach has significantly advanced the field by removing the need for complex IR detectors, which are often limited by high thermal noise and technical challenges \cite{ma2023eliminating,gilaberte2021video}. Recent advancements have focused on optimizing the resolution and imaging efficiency of QIUP systems \cite{fuenzalida2023experimental}. Theoretical and experimental studies have systematically analyzed key parameters such as wavelength, pump waist, and spectral properties that influence image quality and system performance \cite{viswanathan2021resolution,gilaberte2023experimental}. These developments have established a robust foundation for practical implementations, with demonstrated capabilities in extracting IR spectral fingerprints of biological samples, thereby opening new avenues for non-invasive biomedical imaging and material characterization \cite{buzas2020biological,pearce2023practical,kviatkovsky2020microscopy}.\\

A key advancement in quantum imaging with undetected photons (QIUP) is the integration of polarization tomography, which allows researchers to probe how a sample modifies the polarization state of light. By systematically analyzing these polarization changes, it becomes possible to extract detailed information about the sample’s internal structure and composition. Recent studies have shown that the quantum state of the undetected idler photon can be reconstructed through measurements on its entangled signal photon, enabling indirect yet precise characterization \cite{fuenzalida2024quantum,lahiri2021characterizing}. New frameworks, such as visibility Stokes parameters, have further established a reliable foundation for quantum state reconstruction in QIUP systems \cite{kysela2024visibility}. Recent advances have demonstrated that nonlinear interferometers with undetected light enable simultaneous sensing of birefringence and diattenuation, motivating efficient and general polarization tomography protocols for complex samples \cite{oglialoro2025sensing}. \\

In this work, we present a practical and efficient method for polarization tomography using a simplified QIUP setup. By employing polarization-selective nonlinear crystals and a rotatable half-wave plate, we control and measure the horizontal and vertical polarization components, enabling accurate extraction of the sample’s Jones matrix. The protocol first reconstructs the Jones matrix from H/V basis measurements and then refines it with additional probe states for improved reliability and precision. This two-step approach minimizes hardware requirements, reduces ambiguity in parameter estimation, and supports straightforward analysis. The method is readily extendable to arbitrary polarization states and includes complete mathematical derivations for clarity and reproducibility. \\

\section{Mathematical Section}
\begin{figure}[h]
    \centering
    \includegraphics[width=1\linewidth]{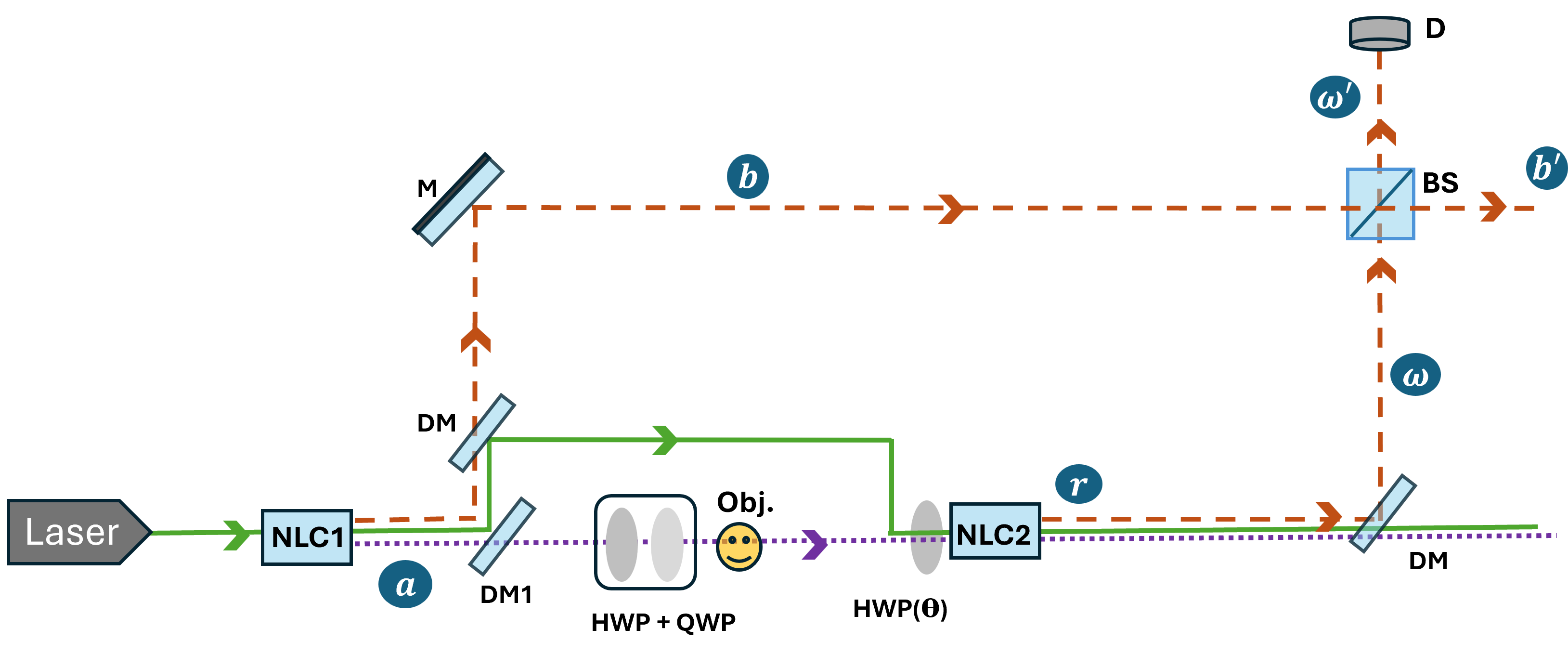}
    \caption{Scheme to explore polarization in QIUP using an additional HWP. Notations: NLC Non-linear crystal, DM Dichroic mirror, HWP Half-wave plate, QWP Quarter-wave plate, BS Beam splitter, M Mirror, D Detector}
    \label{fig:enter-label}
\end{figure}
To begin, we consider the wavefunction of a system comprising two crystals, each generating signal-idler photon pairs in different spatial modes, prior to the introduction of any optical elements.
\begin{equation}
    \ket{\psi_{initial}}= b_1 \ket{V_{I1}}_a \ket{V_{S1}}_a + b_2 \exp{(\iota \zeta)} \ket{V_{I2}}_r \ket{V_{S2}}_r
\end{equation}
Here, $b_1$ and $b_2$ are the generation factors for the two crystals, introduced to balance the contributions from each crystal. The term $\zeta$ is the interferometric phase difference between the photons of the two crystals.
The action of a first dichroic mirror (DM1) is to separate the signal photons emitted from crystal NLC1.
\begin{equation}
    \ket{V_{S1}}_a \rightarrow \ket{V_{S1}}_b
\end{equation}
A controllable polarization transformation of idler 1 photons is implemented using a half-wave plate (HWP) followed by a quarter-wave plate (QWP). 
\begin{equation}
    \ket{V_{I1}}_a \rightarrow \alpha_1 \ket{H_{I1}}_a + \beta_1 \exp{(\iota \gamma)} \ket{V_{I1}}_a
\end{equation}
The state is modified as 
\begin{equation}
\label{SE}
    \ket{\psi}= b_1 (\alpha_1 \ket{H_{I1}}_a + \beta_1 \exp{(\iota \gamma)} \ket{V_{I1}}_a) \ket{V_{S1}}_b + b_2 \exp{(\iota \zeta)} \ket{V_{I2}}_r \ket{V_{S2}}_r .
\end{equation}

\subsection{Without Object}
The half-wave plate (HWP) is analyzed for two fast-axis orientations,$\theta = 0^o$ and $\theta = 45^o$, treated as distinct cases. These specific angles satisfy the indistinguishability condition for vertical and horizontal polarizations, respectively. When the indistinguishability condition is fulfilled for a given polarization, it enables us to extract information about that particular polarization state of the idler photon.
\begin{equation}
\label{HWP}
    \begin{split}
     & \theta = 0^o; \hat{U} \ket{H_{I1}}_a \rightarrow \ket{H_{I1}}_a, \hat{U} \ket{V_{I1}}_a \rightarrow - \ket{V_{I1}}_a , \\   & \theta = 45^o; \hat{U} \ket{H_{I1}}_a \rightarrow -\ket{V_{I1}}_a, \hat{U} \ket{V_{I1}}_a \rightarrow - \ket{H_{I1}}_a . \\  
    \end{split}
\end{equation}
\subsubsection*{Case I: $\theta=0^o$}
After the action of HWP on the idler photons from crystal 1, the state of the system is
\begin{equation}
\label{sim-1}
    \begin{split}
        \ket{\psi} = & b_1 \Big(\alpha_1 \ket{H_{I1}}_a -\beta_1 \exp{(\iota \gamma)}  \ket{V_{I1}}_a\Big) \ket{V_{S1}}_b + b_2 \exp{(\iota \zeta)} \ket{V_{I2}}_r \ket{V_{S2}}_r . \\
    \end{split}
\end{equation}
The idler photons from Crystal 1 arrive just before NLC2, after incurring losses from various optical elements. We account for these losses by modeling a beam splitter, which transmits with amplitude T into path $r$ and reflects with amplitude R into auxiliary path $x$ as following
\begin{equation}
    \begin{split}
        & \ket{H_{I1}}_a \rightarrow T \ket{H_{I1}}_r + R \ket{H_{I1}}_x ,\\
        & \ket{V_{I1}}_a \rightarrow T \ket{V_{I1}}_r + R \ket{V_{I1}}_x . \\
    \end{split}
\end{equation}
Modified state is
\begin{equation}
    \begin{split}
        \ket{\psi} = & b_1 T \Big(\alpha_1 \ket{H_{I1}}_r - \beta_1 \exp{(\iota \gamma)}  \ket{V_{I1}}_r\Big) \ket{V_{S1}}_b + b_2 \exp{(\iota \zeta)} \ket{V_{I2}}_r \ket{V_{S2}}_r \\
        & + b_1 R \Big(\alpha_1 \ket{H_{I1}}_x - \beta_1 \exp{(\iota \gamma)} \ket{V_{I}}_x\Big) \ket{V_{S1}}_b . \\
    \end{split}
\end{equation}
Idler photon indistinguishability is preserved only for those photons that are generated in identical polarization states at the source \cite{arya2025impact}. In the current configuration, where sources emit solely vertically polarized photon pairs, indistinguishability and thus quantum interference applies strictly to the vertical polarization channel. 
\begin{equation}
    \begin{split}
    &   \ket{V_{I1}}_r \rightarrow \ket{V_{I}}_r \\
  &   \ket{V_{I2}}_r \rightarrow \ket{V_{I}}_r \\
  &  \ket{V_{S1}}_{b} \rightarrow \ket{V_{S}}_{b}  \\
  &   \ket{V_{S2}}_{r} \rightarrow \ket{V_{S}}_{r}  \\
    \end{split}
\end{equation}
Therefore, the state takes the form
\begin{equation}
    \begin{split}
        \ket{\psi} = & b_1 T \Big(\alpha_1 \ket{H_{I1}}_r - \beta_1 \exp{(\iota \gamma)}  \ket{V_{I}}_r\Big) \ket{V_{S1}}_b + b_2 \exp{(\iota \zeta)} \ket{V_{I}}_r \ket{V_{S}}_r \\
        & + b_1 R \Big(\alpha_1 \ket{H_{I1}}_x - \beta_1 \exp{(\iota \gamma)} \ket{V_{I}}_x\Big) \\
    \end{split}
\end{equation}
The DM placed afterwards separates the signal photons from path $r$ and direct them towards path $\omega$ as $\ket{V_S}_r \rightarrow \ket{V_S}_{\omega}$. The action of final beam splitter on the photons coming from paths $b$ and $\omega$ into paths $b^{'}$ and $\omega^{'}$ \cite{gerry2023introductory}, and final state is
\begin{equation}
    \begin{split}
        \ket{\psi_{final}} = &  b_1 T \Big(\alpha_1 \ket{H_{I1}}_r - \beta_1 \exp{(\iota \gamma)} \ket{V_{I}}_r\Big) \Big(\frac{\ket{V_s}_{b^{'}} + \iota \ket{V_s}_{\omega^{'}}}{\sqrt{2}}\Big) + b_2 \exp{(\iota \zeta)} \ket{V_I}_r \Big(\frac{\ket{V_s}_{\omega^{'}} + \iota \ket{V_s}_{b^{'}}}{\sqrt{2}}\Big) \\
        & + b_1 R \Big(\alpha_1 \ket{H_{I1}}_x +\beta_1 \exp{(\iota \gamma)} \ket{V_{I}}_x \Big) \Big(\frac{\ket{V_s}_{b^{'}} + \iota \ket{V_s}_{\omega^{'}}}{\sqrt{2}}\Big). \\
    \end{split}
\end{equation}
Vertical counts in path $\omega^{'}$ can be determined as
\begin{equation}
\begin{split}
 \langle N_V \rangle & = \bra{\psi_{final}}\hat{a}^{\dagger}_{V_{\omega^{'}}} \hat{a}_{V_{\omega^{'}}} \ket{\psi_{final}} \\
 & = \frac{b_1^2}{2} + \frac{b_2^2}{2} - b_1 b_2 T \beta_1   \sin{[\gamma - \zeta]}  \\
 \end{split}
\end{equation} 
The visibility using the counts is defined as
\begin{equation}
    \nu = \frac{\langle N_V \rangle_{max} - \langle N_V \rangle_{min}}{\langle N_V \rangle_{max} + \langle N_V \rangle_{min}},
\end{equation}
and determined for our case is
\begin{equation}
    \nu_{\theta=0} = \frac{2 b_1 b_2}{b_1^2 + b_2^2} T \beta_1 .
\end{equation}

For $b_1$ = $b_2$ = $1/\sqrt{2}$, equal contribution by both crystals, the visibility takes the form 
\begin{equation}
\label{vbeta}
    \nu_{\theta=0} =  T \beta_1 
\end{equation}
\subsubsection*{Case II: $\theta=45^o$}
The HWP action (given in Eq.\eqref{HWP}) would modify the state (analogous to Eq.\eqref{sim-1}) in the following way
\begin{equation}
    \begin{split}
        \ket{\psi} = & b_1 \Big(-\alpha_1 \ket{V_{I1}}_a -\beta_1 \exp{(\iota \gamma)}  \ket{H_{I1}}_a\Big) \ket{V_{S1}}_b + b_2 \exp{(\iota \zeta)} \ket{V_{I2}}_r \ket{V_{S2}}_r \\
    \end{split}
\end{equation}
Proceeding in the same way as in the previous case, we arrive at the final expression for the vertical counts as:
\begin{equation}
     \langle N_V \rangle = \frac{b_1^2}{2} + \frac{b_2^2}{2} - b_1 b_2 T \alpha_1   \sin{[ \zeta]}. 
\end{equation}
Again for the ideal case ($b_1$ = $b_2$), the visibility is evaluated as
\begin{equation}
\label{valpha}
    \nu_{\theta=45} =  T \alpha_1 
\end{equation}
The state of the idler photons can be inferred from the corresponding density matrix elements, which are obtained through visibilities measured in the interference patterns (Eqs.\eqref{valpha} and \eqref{vbeta}). To begin with, the transmission amplitude T is determined by preparing the idler in a known polarization state either horizontal ($\alpha$ = 1) or vertical ( $\beta$ = 1) and measuring the visibility. \\
For a general polarization state, the interference pattern is recorded by varying the phase difference ($\zeta$) for $\theta = 0^o$ and $\theta = 45^o$. The resulting visibility allows extraction of the relative amplitudes $\alpha$ and $\beta$. Finally, the relative phase ($\gamma$) between the horizontal and vertical components is deduced from the difference of the fringe patterns, thereby providing full information about the input idler state.

\subsection{With Object}
Resume from Eq.\eqref{SE}, after obtaining the desired state of polarization using HWP and QWP combination, the idler photons from NLC1 interacted with the object as following
\begin{equation}
    \begin{split}
      &  \hat{O} \ket{H_{I1}}_a \rightarrow O_{HH} \ket{H_{I1}}_a + O_{VH} \ket{V_{I1}}_a \\
           &  \hat{O} \ket{V_{I1}}_a \rightarrow O_{HV} \ket{H_{I1}}_a + O_{VV} \ket{V_{I1}}_a \\
    \end{split}
\end{equation}
Therefore, the state takes the form
\begin{equation}
    \ket{\psi} = b_1 \Big(A^{''} \ket{H_{I1}}_a + B^{''} \ket{V_{I1}}_a\Big) \ket{V_{S1}}_b + b_2 \exp{(\iota \zeta)} \ket{V_{I2}} \ket{V_{S2}}_r
\end{equation}
where $A^{''}$ and $B^{''}$ are given below
\begin{equation}
    \begin{split}
        & A^{''} = \alpha_1 O_{HH} + \beta_1 \exp{(\iota \gamma)} O_{HV} \\
        & B^{''} = \alpha_1 O_{VH} + \beta_1 \exp{(\iota \gamma)} O_{VV}
    \end{split}
\end{equation}

\subsubsection*{Case I: $\theta=0^o$}
The state of the system by the action of HWP as given in Eq.(\ref{HWP}) is
\begin{equation}
    \ket{\psi} = b_1 \Big(A^{''} \ket{H_{I1}}_a -B^{''} \ket{V_{I1}}_a\Big) \ket{V_{S1}}_b + b_2 \exp{(\iota \zeta)} \ket{V_{I2}} \ket{V_{S2}}_r.
\end{equation}
Following the same procedure outlined earlier, we arrive at the final state of the system.
\begin{equation}
    \begin{split}
        \ket{\psi_{final}} = & b_1 T \Big(A^{''} \ket{H_{I1}}_r - B^{''} \ket{V_{I}}_r \Big)\Big(\frac{\ket{V_s}_{b^{'}} + \iota \ket{V_s}_{\omega^{'}}}{\sqrt{2}}\Big) + b_2 \exp{(\iota \zeta)} \ket{V_I}_r \Big(\frac{\ket{V_s}_{\omega^{'}} + \iota \ket{V_s}_{b^{'}}}{\sqrt{2}}\Big) \\
      & + b_1 R \Big(A^{''} \ket{H_{I1}}_x - B^{''} \ket{V_{I}}_x \Big) \Big(\frac{\ket{V_s}_{b^{'}} + \iota \ket{V_s}_{\omega^{'}}}{\sqrt{2}}\Big) \\  
    \end{split}
\end{equation}
Vertical counts in path $\omega^{'}$ are determined as
\begin{equation}
 \langle N_V \rangle= \bra{\psi_{final}}\hat{a}^{\dagger}_{V_{\omega^{'}}} \hat{a}_{V_{\omega^{'}}} \ket{\psi_{final}}
\end{equation}
\begin{equation}
     \langle N_V \rangle = \frac{b_1^2}{2}\big(|A^{''}|^2 + |B^{''}|^2\big) + \frac{b_2^2}{2} - \frac{\iota b_1 b_2 T }{2} \big( B^{''} \exp{(-\iota \zeta)} -  B^{{''}^{*}} \exp{(\iota \zeta)} \big)
\end{equation}
If we further consider the object matrix in the following form \cite{kilchoer2019determining},
\begin{equation}
\hat{O} = \begin{pmatrix}
    O_{HH} & O_{HV} \\
    O_{VH} & O_{VV}  
\end{pmatrix} =
    \begin{pmatrix}
        \tau_H \exp{(\iota \phi_H)} & \kappa \exp{(\iota \xi)} \\
        - \kappa \exp{(-\iota \xi)} & \tau_V \exp{(\iota \phi_V)}
    \end{pmatrix}
\end{equation}.
Here, $\tau_H$ and $\tau_V$ are transmission amplitude coefficients, and $\phi_H$ and $\phi_V$ are the phases corresponding to the horizontal and vertical components, respectively. The symbols $\kappa$ and $\zeta$ denote the coupling strength and phase between the horizontal and vertical components.
\begin{equation}
  \begin{split}
        \langle N_V \rangle_{\theta=0} = & \frac{b_2^2}{2} + \frac{b_1^2}{2} \Big( \kappa^2 + \big(\alpha_1^2 \tau_H^2 + \beta_1^2 \tau_V^2\big) + 2 \alpha_1 \beta_1 \kappa \big( \tau_H \cos{[\phi_H -\gamma - \xi]} - \tau_V \cos{[\phi_V +\gamma + \xi]}\big) \Big) \\
        & + b_1 b_2 T \Big( -\alpha_1 \kappa \sin{[\xi+ \zeta]} + \beta_1 \tau_V \sin[\phi_V + \gamma - \zeta] \Big)
  \end{split}
\end{equation}

\subsubsection*{Case I: $\theta=45^o$}
When the state after the HWP changed as
\begin{equation}
    \ket{\psi} = b_1 \Big(- A^{''} \ket{V_{I}}_a -B^{''} \ket{H_{I1}}_a\Big) \ket{V_{S1}}_b + b_2 \ket{V_{I2}} \ket{V_{S2}}_r,
\end{equation}
the resultant vertical counts are obtained as
\begin{equation}
  \begin{split}
        \langle N_V \rangle_{\theta=45} = & \frac{b_2^2}{2} + \frac{b_1^2}{2} \Big( \kappa^2 + \big(\alpha_1^2 \tau_H^2 + \beta_1^2 \tau_V^2\big) + 2 \alpha_1 \beta_1 \kappa \big( \tau_H \cos{[\phi_H -\gamma - \xi]} - \tau_V \cos{[\phi_V +\gamma + \xi]}\big) \Big) \\
        & + b_1 b_2 T \Big( \alpha_1 \tau_H \sin{[\phi_H - \zeta]} + \beta_1 \kappa \sin[\xi + \gamma - \zeta] \Big).
  \end{split}
\end{equation}

\section{Methodology}
In the final stage of our polarization tomography protocol, we emphasize a coherent, scientifically rigorous approach to parameter extraction and validation, ensuring both experimental reliability and theoretical consistency. For the sake of simplification, now we modulate the polarization of the incoming light. Two simple cases are taking either horizontally or vertically polarized light at a moment. 
\begin{equation}
 \begin{split}
        & \Big[\langle N_V \rangle_{\theta=0}\Big]_{\alpha=1} = \frac{b_2^2}{2} + \frac{b_1^2}{2} (\kappa^2 + \tau_H^2) - b_1 b_2 T \kappa \sin[\xi + \zeta] \\
 \end{split}
\end{equation}
\begin{equation}
\label{H}
    \Big[\langle N_V \rangle_{\theta=45}\Big]_{\alpha=1} = \frac{b_2^2}{2} + \frac{b_1^2}{2} (\kappa^2 + \tau_H^2) + b_1 b_2 T \tau_H \sin[\phi_H - \zeta]
\end{equation}
\begin{equation}
\label{V}
 \begin{split}
        & \Big[\langle N_V \rangle_{\theta=0}\Big]_{\beta=1} = \frac{b_2^2}{2} + \frac{b_1^2}{2} (\kappa^2 + \tau_V^2) + b_1 b_2 T \tau_V \sin[\phi_V - \zeta] \\
 \end{split}
\end{equation}
\begin{equation}
    \Big[\langle N_V \rangle_{\theta=45}\Big]_{\beta=1} = \frac{b_2^2}{2} + \frac{b_1^2}{2} (\kappa^2 + \tau_V^2) + b_1 b_2 T \kappa \sin[\xi - \zeta]
\end{equation}
We need to extract the parameters are $\kappa$, $\tau_H$, $\tau_V$, $\phi_H-\phi_V$, $\xi$.  Begin with the ratio of visibilities determined from the above-mentioned vertical counts, 
\begin{equation}
\begin{split}
       & R_1 =\frac{\kappa}{\tau_H} = \frac{\Big[ \nu_{\theta=0}\Big]_{\alpha=1}}{ \Big[ \nu_{\theta=45}\Big]_{\alpha=1}}, \\
       & R_2 = \frac{\tau_V}{\kappa} = \frac{\Big[ \nu_{\theta=0}\Big]_{\beta=1}}{ \Big[ \nu_{\theta=45}\Big]_{\beta=1}}.
\end{split}
\end{equation}
\begin{equation}
  \begin{split}
      &  C_{\alpha} = \frac{b_2^2}{2} + \frac{b_1^2}{2} (\kappa^2 + \frac{\kappa^2}{R_1^2})  \\
      & C_{\beta} = \frac{b_2^2}{2} + \frac{b_1^2}{2} (\kappa^2 + \kappa^2 R_2^2) 
  \end{split}
\end{equation}
The constants $C_{\alpha}$ and $C_{\beta}$ representing the DC term of the vertical counts for $\alpha=1$ and $\beta=1$, respectively. We can determine the $C_{\alpha}$ and $C_{\beta}$ experimentally using the fitting model for the counts in the form $\langle N_V\rangle_{k} = C_{k} + A \sin(\zeta + \phi)$; $k = \alpha, \beta$. Here $A$ denotes the fringe amplitude, $\zeta$ is the scanned interferometric phase, and $\phi$ is the offset phase. This fitting procedure not only yields the baseline counts but also sets the stage for extracting the remaining physical parameters. \\

The value of $\kappa$ determined from two independent equations (e.g., from horizontal and vertical input states) should be nearly equal or within experimental uncertainties. This step ensures the correctness of the experimental data according to the theoretical model. If a significant discrepancy arises, experimental scrutiny is required to identify possible sources of error, such as misalignment, decoherence, mixed states, or a breakdown of the simple Jones matrix assumption. These ratios allow us to express $\tau_H$ and $\tau_V$ directly in terms of the already determined $\kappa$, thus reducing the number of independent unknowns in our model. By anchoring the estimation of these parameters to measurements in the horizontal/vertical basis, we achieve both analytical clarity and experimental efficiency. \\
The phase difference between the horizontal and vertical components ($\phi_H$-$\phi_V$) can be measured through the phase shift of the vertical counts for $\alpha$=1 at $\theta$=$45^o$ and $\beta$=1 at $\theta$=$0^o$ (Eqs.\eqref{H} and \eqref{V}), by fitting the measured interference fringes and extracting the relative phase offset. This approach isolates the relative phase between the horizontal and vertical components, providing direct insight into birefringence or optical path differences within the system. The coupling phase $\xi$, which encapsulates the influence of the off-diagonal elements of the Jones matrix, is extracted by examining the phase of oscillatory terms in the count equations that explicitly depend on $\xi$. By systematically scanning the interferometric phase $\zeta$ and fitting the resulting sinusoidal fringes, we obtain a direct measure of the coupling phase. \\
Although the Jones matrix can, in principle, be reconstructed from the H/V measurements alone, we extend the protocol by performing additional measurements with the idler photon prepared in various polarization states such as diagonal, anti-diagonal, or elliptical states \cite{kilchoer2019determining}. For each additional probe state, the count rate is measured for two angles of HWP ($\theta = 0^o$ and $45^o$). This step yields a set of output equations where the same matrix parameters are involved, but the measurement expressions are distinct, and measurement outcomes will be redundant with respect to the H/V basis solution. o optimally reconcile all available data, we apply a numerical optimization algorithm (e.g., nonlinear least squares), using the H/V-inferred parameter values as initial guesses.  \\
To optimally reconcile all available data, we need to apply a numerical optimization algorithm (e.g., nonlinear least squares), using the H/V-inferred parameter values as initial guesses \cite{brandao2020fast}. This serves two key purposes: \\
Refinement: The optimization process refines parameter values by minimizing the overall error between model predictions and all data, helping to reduce noise and systematic deviations \cite{sugiyama2013precision}. Using H/V parameters as initial guesses ensures faster and more reliable convergence \cite{heinosaari2013quantum}. \\
Validation: By comparing H/V-inferred parameters with data from other probe states, the protocol can identify experimental errors—large discrepancies signal issues that need further investigation \cite{brandao2020fast}. \\
Our methodology extends recent approaches that use variable-polarization probes and interferometric analysis for birefringence and diattenuation sensing \cite{oglialoro2025sensing}, generalizing them into a systematic, optimization-based Jones matrix reconstruction.

\section*{Conclusion}
In summary, we present a resource-efficient protocol for polarization tomography using a QIUP arrangement that requires only vertically polarization-sensitive crystals and accesses horizontal polarization via a half-wave plate, greatly simplifying the experimental setup. The protocol reconstructs the Jones matrix first from H/V measurements, then refines it by incorporating additional probe states and global optimization, starting from the H/V-based estimate. This two-step approach improves accuracy, enables robust error-checking, and reduces ambiguity in parameter estimation while minimizing hardware needs. Our method is scalable and practical, particularly suited for cost-effective and easily aligned quantum polarization tomography in advanced imaging systems.

\section*{Acknowledgment}
The author gratefully acknowledges support from the National Quantum Science and Technology Institute (NQSTI). \\
The author appreciates the valuable conversations and input provided by Dr. Jorge Fuenzalida.

\end{document}